\documentclass[useAMS,usegraphicx,usenatbib]{mn2e}

\usepackage{graphicx}
\usepackage{times}
\usepackage{amssymb}
\usepackage{amsmath}
\newif\ifAMStwofonts
\AMStwofontstrue

\def\asca{{\it ASCA}}
\def\einstein{{\it Einstein}}
\def\chandra{{\it Chandra}}
\def\xmm{{\it XMM-Newton}}

\def\rosat{{\it ROSAT}}

\def\ks{{\rm\thinspace ks}}

\def\sr{{\rm\thinspace sr}}
\def\as{{\rm\thinspace arcsec}}

\def\pcmsq{\hbox{$\rm\thinspace cm^{-2}$}}

\def\kev{{\rm\thinspace keV}}

\def\ctspmspas{\hbox{$\rm\thinspace counts~Ms^{-1}~arcsec^{-2}$}}
\def\ctscmsqperg{\hbox{$\rm\thinspace counts~cm^{2}~erg^{-1}$}}
\def\ergpcmsqps{\hbox{$\rm\thinspace erg~cm^{-2}~s^{-1}$}}

\def\kevpcmsqpspsrpkev{\hbox{$\rm\thinspace keV~cm^{-2}~s^{-1}~sr^{-1}~keV^{-1}$}}

\title{The (un)resolved X-ray background in the Lockman Hole}
\author[M. A. Worsley et al.]
{\parbox[]{6.in}
{M.~A. Worsley,$^{1}$\thanks{E-mail: maw@ast.cam.ac.uk} A.~C. Fabian$,^{1}$ X.~Barcons,$^{2}$ S.~Mateos,$^{2}$ G.~Hasinger$^{3}$ and H.~Brunner$^{3}$}\\\\
\footnotesize
$^{1}$Institute of Astronomy, Madingley Road, Cambridge CB3 0HA\\
$^{2}$Instituto de F\'\i sica de Cantabria (CSIC-UC), 39005 Santander, Spain\\
$^{3}$Max-Planck-Institut f\"ur extraterrestrische Physik, Postfach 1319, D-85740 Garching, Germany\\
}

\begin{document}
\maketitle

\label{firstpage}

\begin{abstract}
Most of the soft and a growing fraction of the harder X-ray background has been resolved into emission from point sources, yet the resolved fraction above $7\kev$ has only been poorly constrained. We use $\sim700\ks$ of \xmm\ observations of the Lockman Hole and a photometric approach to estimate the total flux attributable to resolved sources in a number of different energy bands. We find the resolved fraction of the X-ray background to be $\sim90$ per cent below $2\kev$ but it decreases rapidly at higher energies with the resolved fraction above $\sim7\kev$ being only $\sim50$ per cent. The integrated X-ray spectrum from detected sources has a slope of $\Gamma\sim1.75$, much softer than the $\Gamma=1.4$ of the total background spectrum. The unresolved background component has the spectral signature of highly obscured AGN.
\end{abstract}

\begin{keywords}
surveys -- galaxies: active -- quasars: general -- cosmology: diffuse radiation -- X-rays: galaxies -- X-rays: diffuse background
\end{keywords}

\section{Introduction}

The cosmic X-ray background (XRB) extends from $\sim0.1$ to several hundred $\kev$. At the lowest energies ($\lesssim0.25\kev$) the dominant contribution is the thermal emission from hot gas in the local bubble and the Galaxy. Above $1\kev$ it is mainly extragalactic in origin and is well-described by a power-law with a spectral slope of $\Gamma=1.4$. The spectrum peaks in $\nu I_{\nu}$ at $\sim30\kev$ before decreasing exponentially at higher energies. 

It is now clear that the extragalactic XRB is due to the integrated emission of point sources, principally Active Galactic Nuclei (AGN), along with some thermal emission from galaxy clusters and starbursts (that contribute at soft X-ray energies). Since the early years of the \einstein\ observatory \citep{maccacaro91}, a larger and larger fraction of the XRB has been resolved into discrete objects. Most ($\sim70-80$ per cent) of the $0.5-2\kev$ background was resolved by \rosat\ \citep{hasinger98}. Using \asca, \citet{ueda98} could account for $\sim30$ per cent of the $2-10\kev$ background whilst an early \chandra\ survey of the SSA13 field resolved some $\sim60-80$ per cent \citep{mushotzky00}.

The \chandra\ Deep Fields North and South have now been able to resolve some $\sim70-90$ per cent of the hard XRB \citep[CDF-N;][]{hornschemeier00,hornschemeier01,brandt01_2,brandt01_1,miyaji02,alexander03}, \citep[CDF-S;][]{campana01,tozzi01,giacconi01,giacconi02,rosati02,moretti02}. Significant uncertainty in this fraction results from field-to-field variations.

With the first deep X-ray survey conducted by the \xmm\ observatory \citet{hasinger01} resolved $\sim60$ percent of the $5-10\kev$ hard XRB in the Lockman Hole. This field is one of the most well studied regions of the sky and has been the target of many different multi-wavelength observations. It has an unusually low level of neutral Galactic absorption, only $\sim5\times10^{19}\pcmsq$ compared to the typical $2-3\times10^{20}\pcmsq$ at high latitudes \citep{lockman86,elvis94}. This makes it an excellent window through which to observe distant sources at X-ray energies. 

The $\sim100\ks$ of \xmm\ data used in the \citet{hasinger01} work has now been extended to $\sim700\ks$. The large accumulated observation time; the low level of Galactic absorption; and, most importantly, the large effective area of \xmm\ at energies above $7\kev$ make the data a unique resource. Whilst other studies will focus on source identification and spectra, the aim here is to produce a robust result for the resolved fraction of the XRB, as a function of energy, by summing broad-band fluxes for the detected sources.

Recent synthesis models \citep[e.g.][]{gilli01,gandhi03,ueda03} suggest there may be an unresolved population of highly obscured, faint AGN. Such objects would have negligible flux in the softer energy bands with the population only presenting itself as a missing fraction in the resolved flux at higher energies. \citet{moretti03} compile $\rm{log}\thinspace N-\rm{log}\thinspace S$ distributions and integrate these to determine the resolved fluxes to be $\sim94.3^{+7.0}_{-6.7}$ and $\sim88.8^{+7.8}_{-6.6}$ per cent in the $0.5-2\kev$ and hard $2-10\kev$ bands respectively. Using the recent XRB intensity measurement by \citet{deluca03}, with a higher value of the background normalisation, reduces the hard band resolved fraction to only $\sim78$ per cent. $100$ per cent resolution is not necessarily expected, with maybe $\sim1$ per cent in the hard band resulting from truly diffuse emission \citep{soltan03}. This is not enough, however, to make-up the significant shortfall in the hard band resolved fraction.

Figures based on the broad $2-10\kev$ band can be deceptive since the sensitivity of most instruments in the $7-10\kev$ range is poor, e.g. the effective area of \chandra\ is negligible above $7\kev$; however, \xmm\ remains sensitive (at least in the PN camera) to $\sim12\kev$ and by calculating the resolved fraction separately in a number of finer bands we can analyse much more carefully the behaviour of the fraction with energy.

\section{Method}     

\subsection{Observations and data reduction}

\begin{table}
\centering
\caption{Summary of the \xmm\ Lockman Hole observations. `Tn', `M' and `Tk' indicate the thin, medium and thick filters respectively.}
\label{obs}
\begin{tabular}{llcccccccc}
\hline
Rev. & Date & \multicolumn{3}{c}{Good Exposure time (ks) / filter} \\
     &      & PN    & MOS-1 & MOS-2 \\
\hline
70   & 2000 Mar 27   & 33.4 / Tn  &  33.7 / Tn  &  34.3 / Tk  \\         
71   & 2000 Mar 29   & 31.7 / Tn  &  42.4 / Tk  &  33.9 / Tn  \\
73   & 2000 May 02   & 14.4 / Tn  &  13.8 / Tn  &  13.8 / Tk  \\
74   & 2000 May 05   &  5.0 / Tn  &   5.6 / Tn  &   8.1 / Tk  \\
81   & 2000 May 19   & 27.9 / Tn  &  26.8 / Tn  &  36.2 / Tk  \\
345  & 2001 Oct 27   & 24.3 / M   &  40.6 / M   &  37.1 / M   \\
349  & 2001 Nov 04   & 30.6 / M   &  35.3 / M   &  34.1 / M   \\
522  & 2002 Oct 15   & 55.1 / M   &  79.4 / M   &  81.2 / M   \\
523  & 2002 Oct 17   & 46.4 / M   &  55.5 / M   &  56.2 / M   \\
524  & 2002 Oct 19   & 49.9 / M   &  55.4 / M   &  57.2 / M   \\
525  & 2002 Oct 21   & 61.6 / M   &  78.5 / M   &  78.9 / M   \\
526  & 2002 Oct 23   & 25.7 / M   &  45.5 / M   &  52.4 / M   \\
527  & 2002 Oct 25   & 22.8 / M   &  30.1 / M   &  34.1 / M   \\
528  & 2002 Oct 27   & 13.7 / M   &  27.7 / M   &  33.0 / M   \\
544  & 2002 Nov 27   & 67.8 / M   & 104.1 / M   & 103.2 / M   \\
547  & 2002 Dec 04   & 88.9 / M   &  98.0 / M   &  97.7 / M   \\
548  & 2002 Dec 06   & 72.1 / M   &  86.1 / M   &  86.5 / M   \\
\hline
\multicolumn{2}{r}{Total time (ks):} & 681        & 858         & 878         \\ 
\hline
\end{tabular}
\end{table}

The Lockman Hole has been observed 17 times by \xmm\ during the PV, AO-1 and AO-2 phases of the mission, with total good exposure times of $681$, $858$ and $878\ks$ in the PN, MOS-1 and MOS-2 instruments respectively. Table~\ref{obs} gives the details of the individual observations. Most of the pointings are centred in the same direction, with some variation to overcome CCD chip gaps. EPIC data reduction was performed with the Scientific Analysis Software (\textsc{SAS}) \textsc{v}5.4.1. The raw event files were cleaned for background flares, hot columns and bad pixels, including the standard low-energy filtering used in the processing pipeline. 

\subsection{Source detection}

The cleaned events from each observation and instrument were processed separately and in five different energy bands; $0.2-0.5$, $0.5-2$, $2-4.5$, $4.5-7.5$ and $7.5-12\kev$. Source detection for each instrument, observation and band followed a similar procedure to that described in the \xmm\ \textsc{SAS} User's Guide (Issue 2.1, section 4.12.3): Sliding-box source detection was carried out using the task \textsc{eboxdetect} and the resulting bright source list used by the task \textsc{esplinemap} to produce approximate background maps. Exposure maps were also generated using \textsc{eexpmap}. These contain the actual effective exposure time as a function of position in an image and incorporate the effects of CCD spatial quantum efficiency variations, filter transmission and mirror vignetting.

The separate images, background maps and exposure maps for each of the 17 observations were then added together resulting in one for each instrument and energy band. Sliding-box source detection was performed again, this time utilising the background information. The detection algorithm simultaneously considered all energy bands in the process although, again, the three instruments were processed separately. The resulting source lists were refined further using the maximum-likelihood point spread function (PSF) fitting task \textsc{emldetect} to find the best-fitting position of each source candidate.

The final source lists for each instrument were verified manually to exclude spurious detections due to residual bad pixels and to resolve special cases (e.g. where a closely spaced pair of sources had been erroneously fitted with a single PSF). A second run of the source characterisation task was used, with source positions fixed, to confirm the results. The source lists for each instrument were then cross-checked -- resulting in a final list of 126 source positions. 

Verification of the source list was obtained by comparison with the independently constructed list by Mateos et al. (in prep.); both lists agree well although there are 30 very faint source candidates (out of a total of 156) which are not contained in our list. These sources have fluxes $\lesssim10^{-15}\ergpcmsqps$ ($0.5-7.5\kev$) and were not consistently detected in our analysis, neither were they clear sources when manually checked. The following photometric analysis produces the same results regardless of their inclusion or not (the effect on the resolved XRB intensity is negligible). 

The brightest sources have fluxes $\sim3-5\times10^{-14}\ergpcmsqps$ (depending on energy band), down to $\sim\rm{few}\times10^{-16}\ergpcmsqps$, in each energy band. The limiting factor to source detection is the rather high level of background noise; in the $0.2-0.5\kev$ band it is $\sim0.4-0.6~(0.1-0.2)~\ctspmspas$ for the PN (MOS) camera; increasing to $\sim1.1-1.3~(0.3-0.4)~\ctspmspas$ for the $7.5-12\kev$ band. 

The level of source-confusion, for the faintest sources, can be quantified as the number of `beams' per detected source; where a beam is taken to be the solid angle of one resolution element \citep[see e.g.][]{hogg2001}. Taking a resolution element to be circular, with a diameter of $\sim15\as$, the number of beams per source is $\sim50$ -- above the level at which confusion becomes important ($\lesssim30$ beams per source; \citealp{hogg2001}). 

\subsection{Source photometry}
\label{photometry}

\begin{figure}
\rotatebox{0}{
\resizebox{!}{0.565\columnwidth}
{\includegraphics{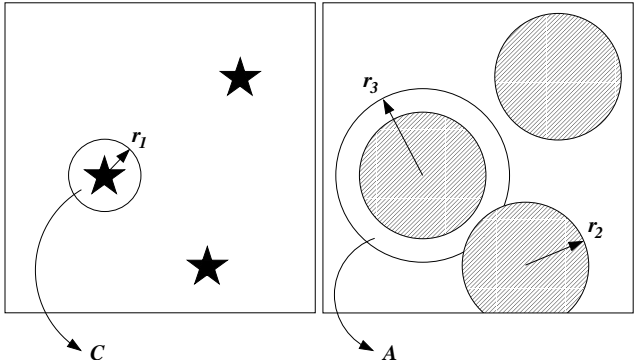}}}
\caption{A schematic representation of the count-rate extraction regions for a source. The left panel shows the exposure-corrected image with the radius $r_1$ aperture used to extract the count-rate $C$. The right panel shows the source-free image where the sources have been masked-out with a radius $r_2$ aperture. The count-rate $A$ is extracted from the surrounding annulus of inner and outer radius $r_2$ and $r_3$ respectively. Portions of the annulus overlapping with the masked-out region of a nearby source are ignored (as shown).}
\label{cutouts}
\end{figure}

A straightforward photometric approach was adopted to obtain the total flux due to resolved sources in each energy band. The three instruments were treated separately to provide robustness in the results and also because of the differences in PSF and background levels. The process was restricted to the sources lying within a $10\rm{'}$ radius centred on $10^{\rm{h}}52^{\rm{m}}41^{\rm{s}}$ $+57^{\rm{\circ}}29\rm{'}00\rm{''}$ (the central portion of the field with the lowest background levels and highest exposure time).

For each instrument and energy band, an exposure-corrected image was made by simply dividing the image by the exposure map. A `source-free' version of each exposure-corrected image was also made by masking out all sources with an energy dependent cut-out radius $r_2$ (see \ref{cutout}). This was done for all sources and also for any rejected source candidates or instrumental artefacts. 

For each source and in each energy band, the total exposure-corrected count-rate $C$ was extracted from within a circular aperture of radius $r_1$. The total exposure-corrected count-rate $A$ was also measured in an annulus surrounding the source (from a radius $r_2$ to $r_3$) but using the source-free image to limit contamination by other nearby sources. Errors in $C$ and $A$ are assumed to be Poissonian. Fig.~\ref{cutouts} is a schematic of the image cut-out aperture and the source-free cut-out annulus for a source. 

The true source count-rate $S$ ($\rm{counts~s^{-1}}$) and the background level $B$ ($\rm{counts~s^{-1}~arcsec^{-2}}$) can be determined from the measured $C$ and $A$ via,

\begin{equation}
S = \frac{QC-PA}{QE(r_1)-P[E(r_3)-E(r_2)]}
\label{eqn:S}
\end{equation}
\begin{equation}
B = \frac{AE(r_1)-C[E(r_3)-E(r_2)]}{QE(r_1)-P[E(r_3)-E(r_2)]}
\label{eqn:B}
\end{equation}
where $P$ and $Q$ ($\rm{arcsec^{2}}$) are the areas of the cut-out aperture and annulus respectively. $E(r)$ is the fraction of the source counts which would fall within a cut-out radius $r$ -- the `encircled energy fraction' (EEF) -- computed by integrating the analytic fit to the PSF given in \xmm\ calibration documents EPIC-MCT-TN-011 (2001 October 8) and EPIC-MCT-TN-012 (2002 June 27). 

The analytical PSF is a radially-symmetric King function, with parameters that vary with photon energy and off-axis angle. As detailed in the calibration documents, the analytical model was constructed by considering radially-averaged PSF profiles and include the effects of the azimuthal distortions seen in the PSF when far off-axis. The models for the three instruments are only well-determined within specific `ranges of application' and have not been verified for larger off-axis angles at higher energies. 

In order to justify the use of the analytical model in the harder energy bands and for sources far off-axis, we tested the analytical fit to the PSF profiles for several bright sources that lie in different parts of the field. The combined images were used and the profiles in each band fitted to the analytical model. An acceptable fit was obtained in each case -- the analytical PSF would appear to be a satisfactory model for all the sources in our field, including those which strictly fall outside `range of application'. 

A further complication in the use of the analytical model in the combined images; these were made by stacking 17 different individual images, not all of which are in the exactly same pointing direction. The PSF of a source is strictly a superposition of PSFs with different off-axis angles. When telescope vignetting is taken into account, the majority of counts ($\sim90$ per cent) are due to data from observations with pointing directions that lie within $\sim1.5\rm{'}$ of the centre of the combined field and the effect is not significant. The aforementioned test of the analytical PSF model was carried out using sources from the combined images and the success of the model in fitting the actual PSF profiles demonstrates that it remains valid for use in the combined data sets.

\subsection{Cut-out radii}
\label{cutout}

\begin{table}
\centering
\caption{The cut-out radii used for the PN and MOS instruments and energy bands, as described in Fig.~\ref{cutouts}.}
\label{cutoutnumbers}
\begin{tabular}{cccccccc}
\hline
Energy band & \multicolumn{7}{c}{Cut-out radii (arcsec)}          \\
(keV)       & \multicolumn{3}{c}{PN}  &  & \multicolumn{3}{c}{MOS} \\
            & $r_1$ & $r_2$ & $r_3$   &  & $r_1$ & $r_2$ & $r_3$   \\
\hline
$0.2-0.5$   & 13    & 40    & 50      &  & 12    & 44    & 54      \\
$0.5-2$     & 15    & 44    & 54      &  & 13    & 36    & 46      \\
$2-4.5$     & 11    & 32    & 42      &  & 11    & 32    & 42      \\
$4.5-7.5$   & 8     & 24    & 34      &  & 7     & 20    & 30      \\
$7.5-12$    & 6     & 16    & 26      &  & 6     & 12    & 22      \\            
\hline
\end{tabular}
\end{table}

Radii $r_2$ (for masking sources in the source-free images) were determined empirically by considering the observed PSFs of a sample of some typical sources (i.e. from different parts of the field, of moderate brightness and comparatively isolated). Counts were plotted as a function of radius away from the centroid position of the source. The mask-out radius was taken to be the radius at which the count-rate (within Poisson error) decreases to the background level. This was performed for each energy band and instrument since both the PSF and background can vary considerably between them.

Source cut-out radii $r_1$ have only a weak influence on results since the extracted source counts are corrected for the fact that the encircled energy fraction within the cut-out radius is not 100 per cent. However, it is important to choose an optimal value to maximise the signal-to-noise -- a small radius will reject too many source photons whilst a large radius will include too many background photons. For a typical source brightness and background level, the optimal cut-out radii were determined for each energy band and instrument by maximising the signal-to-noise in $S$. Table~\ref{cutoutnumbers} gives the values of the different cut-out radii used for the different instruments and energy bands. The encircled energy fraction within the cut-out radii is $\sim45-65$ per cent (depending on energy band and instrument) and so the EEF correction factor is $\sim2.2-1.5$).

The masking radii $r_2$ and the source cut-out radii $r_1$ are much smaller at higher energies for two reasons; firstly, the
core of PSF is narrower because high energy photons are only focused by the innermost shells of the telescope and the reduced number results in less dispersion; secondly, the background level is higher in the harder bands -- meaning the radius at which source counts are dominated by background is smaller.

%There is negligible overlap of the extraction apertures between sources although this remains a small source of error. Contamination of a particular extraction aperture by counts nearby sources (even if the apertures do not overlap) is also negligible. This effect is lessened since there would be similar contamination of the annulus (and the measurement of $A$), a compensating factor since the polluting counts in both the aperture and annulus will contribute towards the background level.

\subsection{Count-rate to flux conversion}

\begin{table}
\centering
\caption{The energy conversion factors used to scale source count-rates to fluxes. Each ECF is a combination of the different values for different filters, weighted according to the exposure time spent in each filter for the particular instrument and energy band.}
\label{ecfs}
\begin{tabular}{cccc}
\hline
Energy band & \multicolumn{3}{c}{Weighted energy conversion factor} \\
(keV)       & \multicolumn{3}{c}{($10^{11}\ctscmsqperg$)} \\
            & PN      & MOS-1   & MOS-2 \\
\hline
$0.2-0.5$   & $7.409$ & $1.261$ & $1.240$ \\
$0.5-2$     & $6.379$ & $1.922$ & $1.903$ \\
$2-4.5$     & $1.910$ & $0.723$ & $0.724$ \\
$4.5-7.5$   & $0.933$ & $0.259$ & $0.271$ \\
$7.5-12$    & $0.234$ & $0.027$ & $0.029$ \\
\hline
\end{tabular}
\end{table}

The final count-rate $S$ for each source in the field was converted to a flux measurement. The `energy conversion factors' (ECFs) were computed for each instrument using the same method used to calculate those for the \xmm\ processing pipeline (R. Saxton, priv. comm.; also see the \xmm\ SSC documents SSC-LUS-TN-0059 versions 1-3).

The ECFs were estimated by taking an on-axis source (for the required instrument and filter) and extracting a spectrum. The appropriate redistribution matrix file (RMF) and ancillary response file (ARF) were generated (using the median observation date for the Lockman Hole data). Note that no encircled energy fraction correction is required in the creation of the ARF: this effect is taken into account during source photometry (see \ref{photometry}). Using the X-ray spectral fitting software \textsc{xspec} the spectrum was fitted with a power-law plus Galactic absorption model (using a power-law slope of $\Gamma=1.4$ and an absorption column of $N_{\rm{H}}=5\times10^{19}\pcmsq$; appropriate to the Lockman Hole). A simulated spectrum for the model was then created (with the \textsc{fakeit} command) but with no simulated noise. The ECF was then found by dividing the simulated count-rate by the integrated model flux in the required energy range. 

The ECFs for different filters were combined, weighted according to the exposure spent in each filter, leaving the ECF values for each instrument and energy band. Table~\ref{ecfs} lists the ECFs used. We place a conservative error of 10 per cent on the calculated ECFs to account for the variation of source spectral shapes around the expected XRB `average' for $\Gamma=1.4$ and $N_{\rm{H}}=5\times10^{19}\pcmsq$, as well as uncertainties introduced by the evolution of RMF/ARFs throughout the observations. 

\subsection{Resolved XRB intensity}

In each energy band the source fluxes were summed to give a total flux. All sources (from the final detection list) were considered in each band so even negative fluxes are included in the sum (if negative values are ignored the resolved fraction is artificially higher although the same trend is seen in the results). The flux errors were added appropriately and combined with the estimated error in the ECF values. A correction was made for Galactic absorption: this is $\sim15$ per cent in the $0.2-0.5\kev$ band and insignificant in the harder bands ($<1$ per cent). The apparent resolved flux was then converted to the resolved XRB intensity over the relevant $10'$ source-inclusion circle.

\section{Results}

\begin{figure}
\rotatebox{270}{
\resizebox{!}{\columnwidth}
{\includegraphics{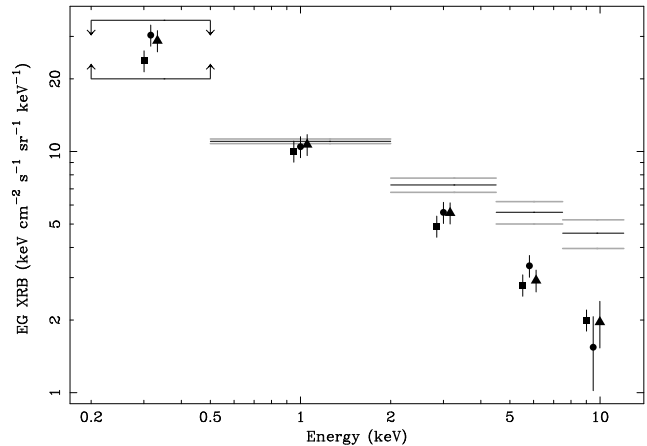}}}
\caption{The extragalactic XRB intensity resolved from detected sources in the energy bands ($0.2-0.5$, $0.5-2$, $2-4.5$, $4.5-7.5$ and $7.5-12\kev$). The values from the three instruments are plotted, for clarity, as points in the centre of each band and offset horizontally with respect to each other. Squares, circles and triangles represent the PN, MOS-1 and MOS-2 instruments respectively. Errors are one sigma. The bars represent the measured values of the total extragalactic XRB intensity; the $0.5-2$, $2-4.5$, $4.5-7.5$ and $7.5-12\kev$ values from \citet{deluca03} with the grey bars indicating estimated one sigma error; the  $0.2-0.5\kev$ data is from \citet{warwick98} and shows the upper and lower bounds of their estimate.}
\label{intensity}
\end{figure}

The resolved XRB intensity for each energy band and instrument in shown in Fig.~\ref{intensity}. In comparison, the most recent total extragalactic XRB intensity measurement \citep{deluca03} is also shown. This work was based on analysis of the $2-8\kev$ XRB - mostly extragalactic in origin at these energies. The extrapolation to the $0.5-2$ and $7.5-12\kev$ bins is acceptable given the assumption that the extragalactic component has the same power-law spectrum at such energies. The extragalactic XRB level is very uncertain at the softest energies (i.e. $0.2-0.5\kev$) and is estimated to lie at $20-35\kevpcmsqpspsrpkev$ by \citet{warwick98}. 

\begin{figure}
\rotatebox{270}{
\resizebox{!}{\columnwidth}
{\includegraphics{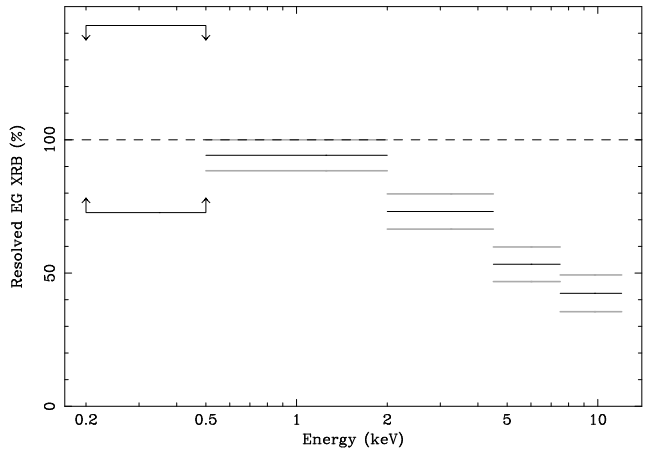}}}
\caption{The fraction of the total extragalactic XRB intensity resolved from detected sources. The combined result from all three instruments is shown, the grey bars indicate one sigma errors for each fraction.}
\label{fraction}
\end{figure}

\begin{figure}
\rotatebox{270}{
\resizebox{!}{\columnwidth}
{\includegraphics{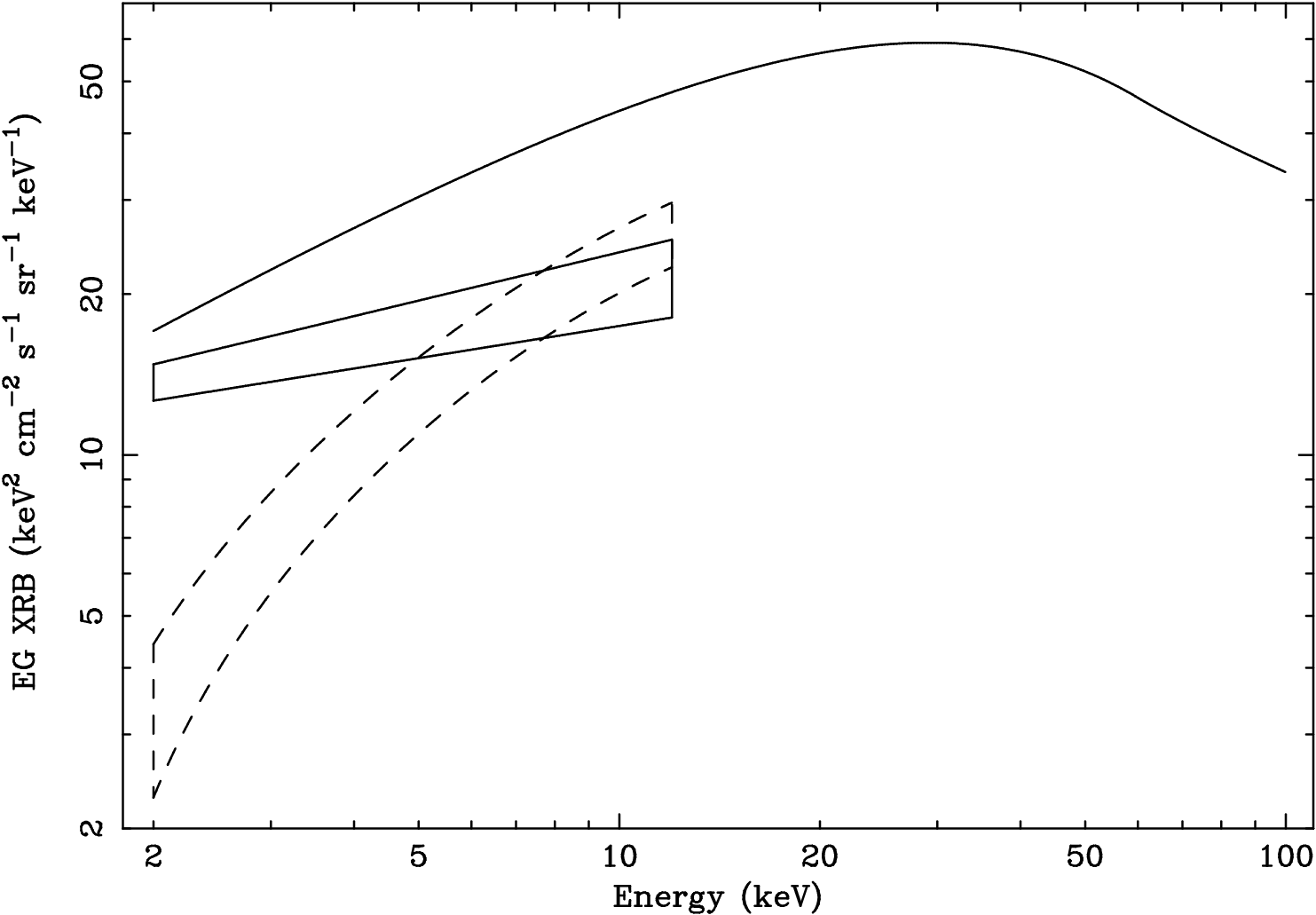}}}
\caption{The total XRB as measured by the HEAO-1 missions \citep{gruber99} but renormalised to the $2-8\kev$ intensity observed by \citet{deluca03} (a scaling factor of $\sim1.4$). The solid region indicates the $\Gamma=1.75$, $N=11\pm0.5\kevpcmsqpspsrpkev$ resolved spectrum and uncertainty. The dashed region indicates the residual unresolved component and uncertainty.}
\label{residual}
\end{figure}

Fig.~\ref{fraction} shows the resolved fraction of the total XRB. At the softest $0.2-0.5\kev$ energies the XRB resolved fraction is consistent with $100$ per cent, although this is rather uncertain given the lack of knowledge of the absolute level. The residual available to group and galaxy formation is small \citep[see e.g.][]{wu01}. The $0.5-2\kev$ soft band resolved fraction is $\sim90$ per cent, consistent with the value found by \citet{moretti03} which includes the (deeper) CDF observations. The total $2-7.5\kev$ hard band resolved fraction is $\sim60-70$ per cent, again slightly less than the $\sim80$ per cent for the CDF hard band \citep{deluca03} as would be expected. 

Our critical finding is the strong downturn in the XRB resolved fraction seen in the individual energy bands, decreasing from $\sim70$ per cent at $2-4.5\kev$ to $\lesssim50$ per cent at energies $>4.5\kev$. The detected sources are failing to account for the hard XRB. A power-law fitted to the intensity resolved by the three instruments (over all bands) has a normalisation of $N=11.5\pm0.5\kevpcmsqpspsrpkev$ (at $1\kev$) and a spectral slope $\Gamma=1.75\pm0.05$. The data however suggest a steepening of the slope with energy, with $\Gamma\sim1.65$ over $0.5-4.5\kev$, 1.77 over $2-7.5\kev$ and $\sim1.80$ over $4.5-12\kev$, all are considerably softer than the total XRB slope of $\Gamma\sim1.4$. Fig.~\ref{residual} shows the total extragalactic XRB spectrum, along with that resolved here and the missing component. Our result agrees with the trend indicated by \citet[][\space Fig.~19]{barger02} who colour-corrected the CDF-N sources assuming a fixed intrinsic spectrum to show that the resolved fraction does not extrapolate to give the observed XRB.

The level of the unresolved component depends upon the Lockman Hole being a representative sample of the sky, as well as the normalisation of the total XRB. The brightest sources considered have fluxes up to $\sim10^{-13}\ergpcmsqps$; there is a non-negligible contribution to the whole-sky XRB by brighter sources and a further correction needs to be made, increasing the resolved component by $\sim10-20$ per cent (estimated from the $\rm{log}\thinspace N-\rm{log}\thinspace S$ curves compiled by \citealt{moretti03}). This effect is only comparable to the potential field-to-field variations on the sky and regardless of any net adjustments to normalisation, the trend toward a lower resolved fraction at harder energies is still robust.

\begin{figure}
\rotatebox{270}{
\resizebox{!}{\columnwidth}
{\includegraphics{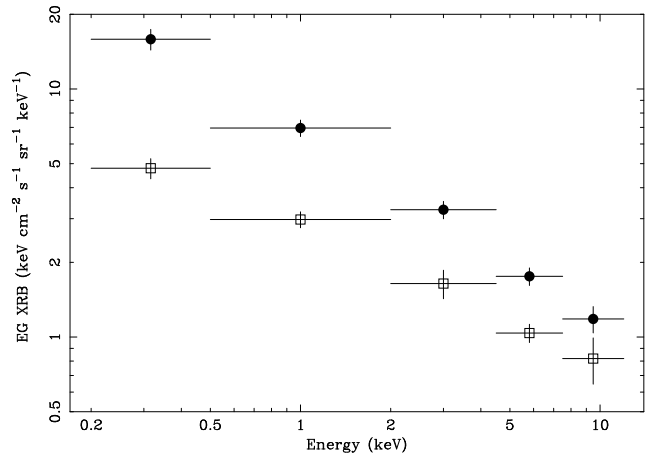}}}
\caption{The resolved XRB intensities from the bright and faint members of the detected source population. The filled circles show the resolved intensity due to the objects with $0.2-12\kev$ fluxes exceeding $2\times10^{-14}\ergpcmsqps$, the open squares show the resolved intensity from the objects fainter than this. The spectrum resolved from faint sources has a spectral slope of $\Gamma\sim1.6$ compared to $\Gamma\sim1.8$. for the bright sources.}
\label{intensity_split}
\end{figure}

Fig.~\ref{intensity_split} again shows the total background level in the five energy bands but the resolved intensity is shown from the faint ($0.2-12\kev$ flux $<2\times10^{-14}\ergpcmsqps$) and bright ($>2\times10^{-14}\ergpcmsqps$) sources separately. The two populations have significantly different spectral slopes. The faint sources are much harder than the bright ones -- the resolved intensity having a spectral slope of $\Gamma\sim1.6$ compared to $\Gamma\sim1.8$ for the bright sources. (power-law normalisations at $1\kev$ are $\sim3.3\kevpcmsqpspsrpkev$ and $\sim8.3\kevpcmsqpspsrpkev$ respectively). 

\section{Discussion}

We have shown that only $\sim50$ per cent of the hard X-ray background above $7\kev$ is resolved and we are missing a considerable population of faint, hard AGN. This is supported by the hardening of the XRB intensity between the spectrum measured from bright sources compared to that of the fainter sources (Fig.~\ref{intensity_split}). The missing component (Fig.~\ref{residual}) has the shape of a heavily absorbed AGN spectrum. Given the negligible X-ray flux, the scattered soft component must be weak, consistent with the emission from almost complete ($4\pi\sr$) coverage by obscuring matter.

\begin{figure}
\rotatebox{270}{
\resizebox{!}{\columnwidth}
{\includegraphics{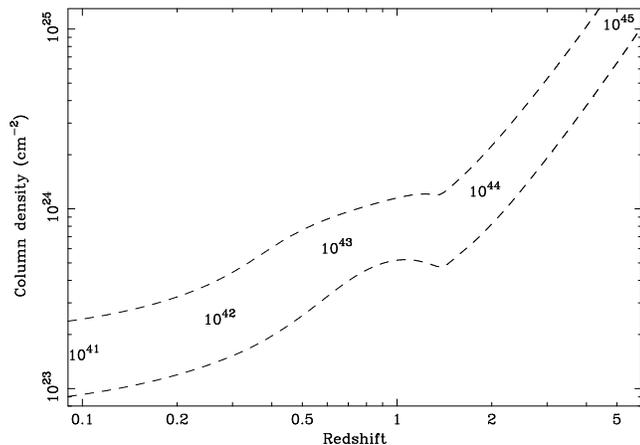}}}
\caption{The column density required to harden an unobscured $\Gamma=2$ spectrum into that of the residual XRB component, as a function of redshift. The residual component (Fig.~\ref{residual}) was approximated to a power-law, the upper and lower bounds plotted are the results for the extremes of the power-law slope for the component ($\Gamma=-0.2$ and $\Gamma=0.9$). The numbers within the plot are the approximate maximum value of the unosbcured, rest-frame, $2-10\kev$ luminosity of the undetected sources ($\rm{erg~s^{-1}}$) in order to remain below the sensitivity limit (for that redshift and column density).}
\label{obscuration}
\end{figure}

Fig.~\ref{obscuration} shows the column density required to harden an intrinsic $\Gamma=2$ emission spectrum into the shape of the faint, hard, unresolved population which has a power-law slope of $\Gamma\sim0.6$ (although note the spectrum is rather curved with extremes of $\sim-0.2$ and $\sim0.9$ at the ends). The column density is greater than $10^{23}\pcmsq$ and Compton-thick objects are required at redshifts $\gtrsim2$. Such AGN are common at low z -- two of the three nearest AGN; NGC\thinspace4945 and the Circinus galaxy, are Compton thick \citep[e.g.][]{matt00} -- few have been found at high redshift \citep{stern02,norman02,crawford03,gandhi04}. Distant ones are difficult to detect individually \citep{fabian02}. The unresolved hard XRB component we have detected has the signature of a substantial obscured AGN population.

\section{Acknowledgments}

Based on observations with \xmm, an ESA science mission with instruments and contributions directly funded by ESA Member States and the USA (NASA). MAW acknowledges support from PPARC and would like to thank Richard Saxton, David Alexander and Franz Bauer for useful discussions. ACF thanks the Royal Society for support. XB acknowledges financial support by the Spanish Ministerio de Ciencia y Tecnolog\'\i a, under project ESP2003-00812.

\bibliographystyle{mnras} %% MNRAS
\bibliography{mn-jour,worsley_08jul04}

\end{document}